\documentclass[a4paper]{spie}
\usepackage[dvips]{graphicx}
\usepackage{epsfig}

\title{Atmospheric refractivity effects on mid-infrared ELT adaptive optics}
\author{Sarah Kendrew\supit{a}, Laurent Jolissaint\supit{a}, Richard J. Mathar\supit{a}, Remko Stuik\supit{a}, Stefan Hippler\supit{b} and Bernhard Brandl\supit{a} \skiplinehalf
    \supit{a}Leiden Observatory, University of Leiden, PO Box 9513, 2300 RA Leiden, Netherlands\\
    \supit{b}Max Planck Institute for Astronomy, K\"{o}nigstuhl 17, 69117 Heidelberg, Germany
    }

\begin{document}
\maketitle

\authorinfo{Send correspondence to: S. Kendrew, Leiden Observatory, kendrew@strw.leidenuniv.nl, tel: +31-71-5278456}

\abstract{We discuss the effect of atmospheric dispersion on the performance of a mid-infrared adaptive optics assisted instrument on an extremely large telescope (ELT). Dispersion and atmospheric chromaticity is generally considered to be negligible in this wavelength regime. It is shown here, however, that with the much-reduced diffraction limit size on an ELT and the need for diffraction-limited performance, refractivity phenomena should be carefully considered in the design and operation of such an instrument. We include an overview of the theory of refractivity, and the influence of infrared resonances caused by the presence of water vapour and other constituents in the atmosphere. `Traditional' atmospheric dispersion is likely to cause a loss of Strehl only at the shortest wavelengths (L-band). A more likely source of error is the difference in wavelengths at which the wavefront is sensed and corrected, leading to pointing offsets between wavefront sensor and science instrument that evolve with time over a long exposure. Infrared radiation is also subject to additional turbulence caused by the presence of water vapour in the atmosphere not seen by visible wavefront sensors, whose effect is poorly understood. We make use of information obtained at radio wavelengths to make a first-order estimate of its effect on the performance of a mid-IR ground-based instrument. The calculations in this paper are performed using parameters from two different sites, one `standard good site' and one `high and dry site' to illustrate the importance of the choice of site for an ELT.}

\bigskip
\section{Mid-infrared adaptive optics in the ELT era}\label{sec:intro}

The next generation of ground-based optical/IR telescopes, the Extremely Large Telescopes (ELTs) will require substantial seeing correction to get the full scientific gain from the large aperture sizes. Scaling today's adaptive optics (AO) technologies to the sizes required for ELTs poses significant challenges: AO systems will be fully integrated into the telescope's operation, either by including a large adaptive mirror in the telescope's main optical path\cite{gilmozzi07}, or by requiring instruments to have built-in AO systems\cite{crampton07}. To provide better sky coverage and correction over a wider field than today's systems, laser guide stars (LGS) will become routinely used alongside natural guide stars, either in single or multiple configurations, thus making possible more advanced AO observing modes such as multi-object AO (MOAO), multi-conjugate AO (MCAO) and laser tomographic AO (LTAO). Both technologically and operationally, this poses some very complex demands.

At mid-IR wavelengths ($\lambda >$ 3 $\mu$m) AO will for the first time be required to observe at or close to the diffraction limit on ground-based telescopes. As the Kolmogorov turbulence power spectrum drops off steeply with increasing wavelength, diffraction limited performance in the mid-IR should be significantly less challenging than at optical wavelengths. Indeed, on today's 8-m-class telescopes, AO becomes obsolete longwards of 17 $\mu$m as observations are intrinsically limited by diffraction rather than seeing. On the scales of an ELT, however, an improvement in resolution of almost an order of magnitude is available in the N-band by correcting for atmospheric seeing.

Despite the AO requirements for mid-IR observations being less strenuous than at visible and near-infrared (NIR) wavelengths, applying AO corrections to mid-IR observations brings its own very specific problems.

\begin{description}
\item[Using NIR-optimised AO systems.] Where ELTs use large adaptive mirrors in the optical path of the telescope, such as in the European ELT (E-ELT), the system will be tailored to NIR observations; this can give rise to chromatic errors when applied in the mid-IR.
\item[Thermal background subtraction.] Subtraction of the thermal background is a crucial part of observing mid-IR radiation from the ground. Traditional chopping methods may not be possible on ELT-sized telescopes because of the large mirror sizes, so alternative methods must be identified or developed\cite{brandl_metis08}. The combination of AO correction and fast field switching observations for background will require careful planning and calibration.
\item[Atmospheric turbulence in the IR.] Infrared radiation is sensitive to the presence of water vapour in the atmosphere, which causes partial or complete opacity of the atmosphere in certain parts of the IR spectrum. In addition, the presence of water vapour is expected to cause increased differential refraction and atmospheric turbulence. These issues are discussed in more detail in the following sections.
\end{description}

This paper discusses a number of problems with carrying out AO-assisted mid-IR observations on an ELT that arise from the dispersion of the refractive index of air, and the sensitivity of IR radiation to the presence of water vapour in the atmosphere. Of particular interest are those effects expected to occur when AO wavefront sensing is performed at a different wavelength than the science observations, as is the likely situation for a mid-IR ELT instrument. A detailed understanding of this chromaticity is essential for assessing the wavefront sensing and dispersion correction needs for a mid-IR ELT instrument designed for (near-) diffraction-limited operation.

Section~\ref{sec:dispersion} introduces the concept of refractivity and how it evolves from visible to mid-IR wavelengths. We describe the formulation we use for calculations of refractivity and atmospheric input parameters. Section~\ref{sec:basic_chromatic} discusses problems arising from `traditional' atmospheric dispersion, i.e. the change in apparent zenith distance of celestial objects from atmospheric refractivity and its likely impact on image quality in the md-IR. It then goes on to describe in more detail the effect of chromaticity on AO correction, for example where wavefront sensing and science wavelengths are different. Section~\ref{sec:wvturb} shows the importance of turbulence caused by water vapour fluctuations on the performance of a mid-IR instrument on an ELT.

The work is being carried out in the context of METIS (the \underline{M}id-infrared \underline{E}-EL\underline{T} \underline{I}mager and \underline{S}pectrograph). A Phase A study for METIS is currently being carried out by an international collaboration of scientists and engineers from the Netherlands, Germany, France, Belgium and the United Kingdom. A detailed description of the project, science case and instrument parameters can be found in Brandl et al.\cite{brandl_metis08}. The AO issues listed above form an important part of the study.

\section{Atmospheric refractivity in the infrared}\label{sec:dispersion}
Atmospheric dispersion has long been known to affect the results of astronomical observations; its effects are commonly described in standard texts\cite{astrophysical}. Several papers\cite{filippenko82, reardon06, arribas99} describe the impact of atmospheric dispersion on particular observation methods or science cases. More recently, atmospheric dispersion has discussed in the context of adaptive optics, where it can contribute significantly to the wavefront error budget\cite{mette06, roe02, helminiak08}. Because of the steep dispersion of the refractive index of air at optical and NIR wavelengths, atmospheric dispersion mostly affects optical/NIR observations. Indeed, the vast majority of available literature only extends its treatment of the phenomenon to NIR wavelengths; a noted exception is found in Hawarden et al.\cite{hawarden06}, where the authors' discussion is extended to the thermal IR regime. However, a thorough understanding of all atmospheric phenomena affecting mid-IR observations is required to achieve diffraction-limited performance on the one hand, and to define any potential dispersion correction needs for the instrument on the other.

To understand chromatic errors introduced by atmospheric dispersion in the infrared, we should revisit the theory of the refractive index of air. This quantity in itself has been the subject of a large volume of work spanning many decades; the availability of large spectral line databases such as HITRAN\cite{hitran} for common atmospheric constituents has enabled a substantial improvement of its accuracy over extended wavelength ranges. The knowledge of the refractive index at IR wavelengths, however, is still lagging behind that for the visible and radio ranges.  Valuable work was published by Hill et al.\cite{hill80, hill86}, and in more recent years by  Mathar\cite{mathar_ao04,mathar07a}.

Two formulae for the refractive index of air are commonly found in the (astronomical) literature. The first is based on early work by Edl\'en\cite{edlen53}, with subsequent revisions by both Edl\'en\cite{edlen66} and other authors in the 1990s\cite{birchdowns93, birchdowns94}. The second, by Ciddor\cite{ciddor96}, is the equation officially accepted by the International Association of Geodesy (IAG). In the equation's validity range to $\lambda=$1.7 $\mu$m, it agrees with the older Edl\'en expression to $1$ part in $10^8$. As the effect of water vapour at visible/NIR wavelengths is smaller, it is often ignored~\cite{bonsch98}. In the infrared, the refractive index of air has contributions from both a slow-varying continuum and strong infrared resonances resulting from absorption by water, CO$_2$ and other atmospheric constituents, giving rise to a complex spectrum that varies significantly with atmospheric temperature, pressure and composition. This makes a simple extrapolation of the refractivity at visible/NIR wavelengths to longer wavelengths unsuitable for precision modelling. Mathar~\cite{mathar_ao04, mathar07a} gives a detailed description of the effect of IR resonances on the refractivity and we use his models thoughout this paper for refractivity calculations. A different method is described by Colavita et al.~\cite{colavita04}, however the authors note that their method is in agreement with that of Mathar.

\begin{figure}
\centering
\includegraphics[width=8cm]{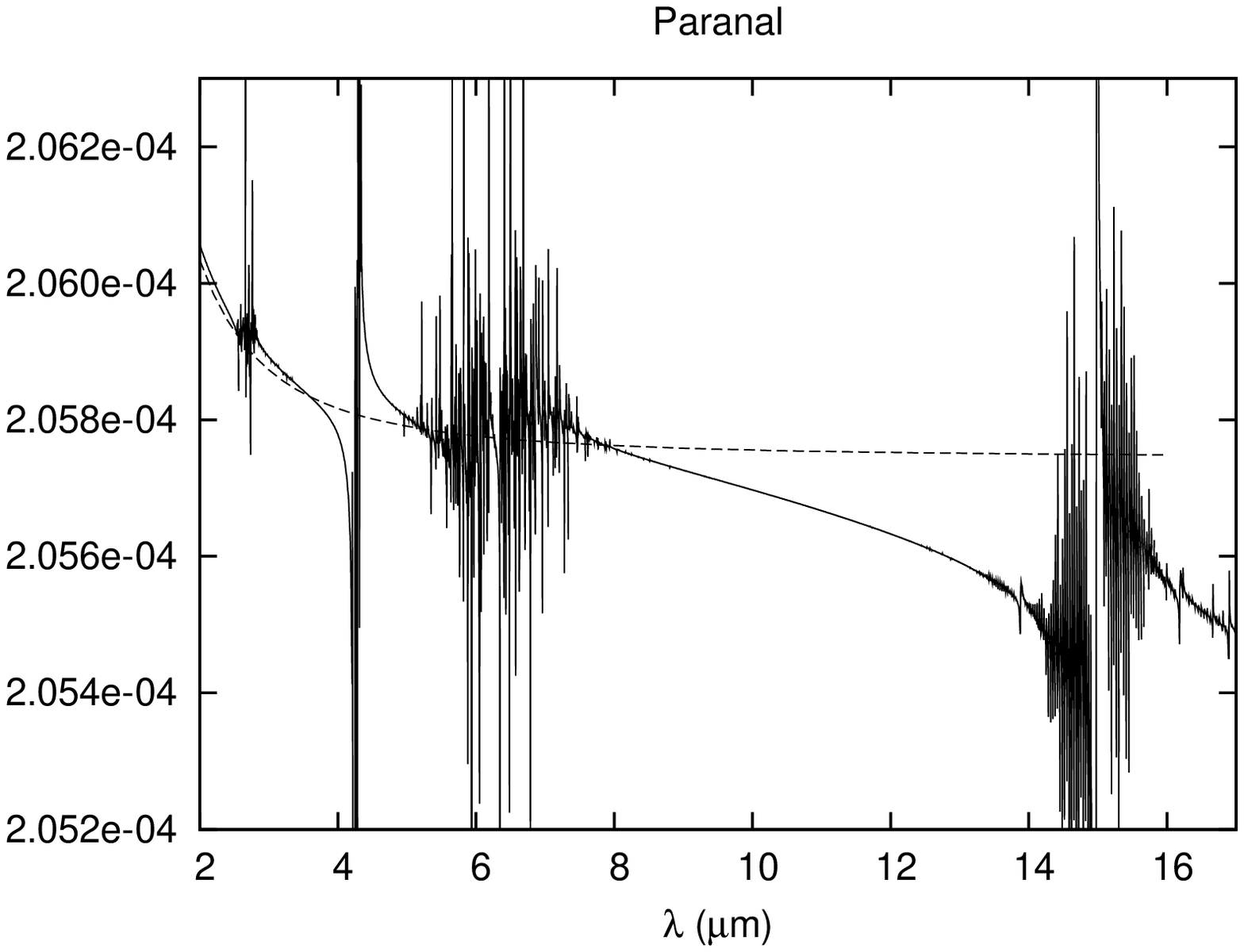}
\includegraphics[width=8cm]{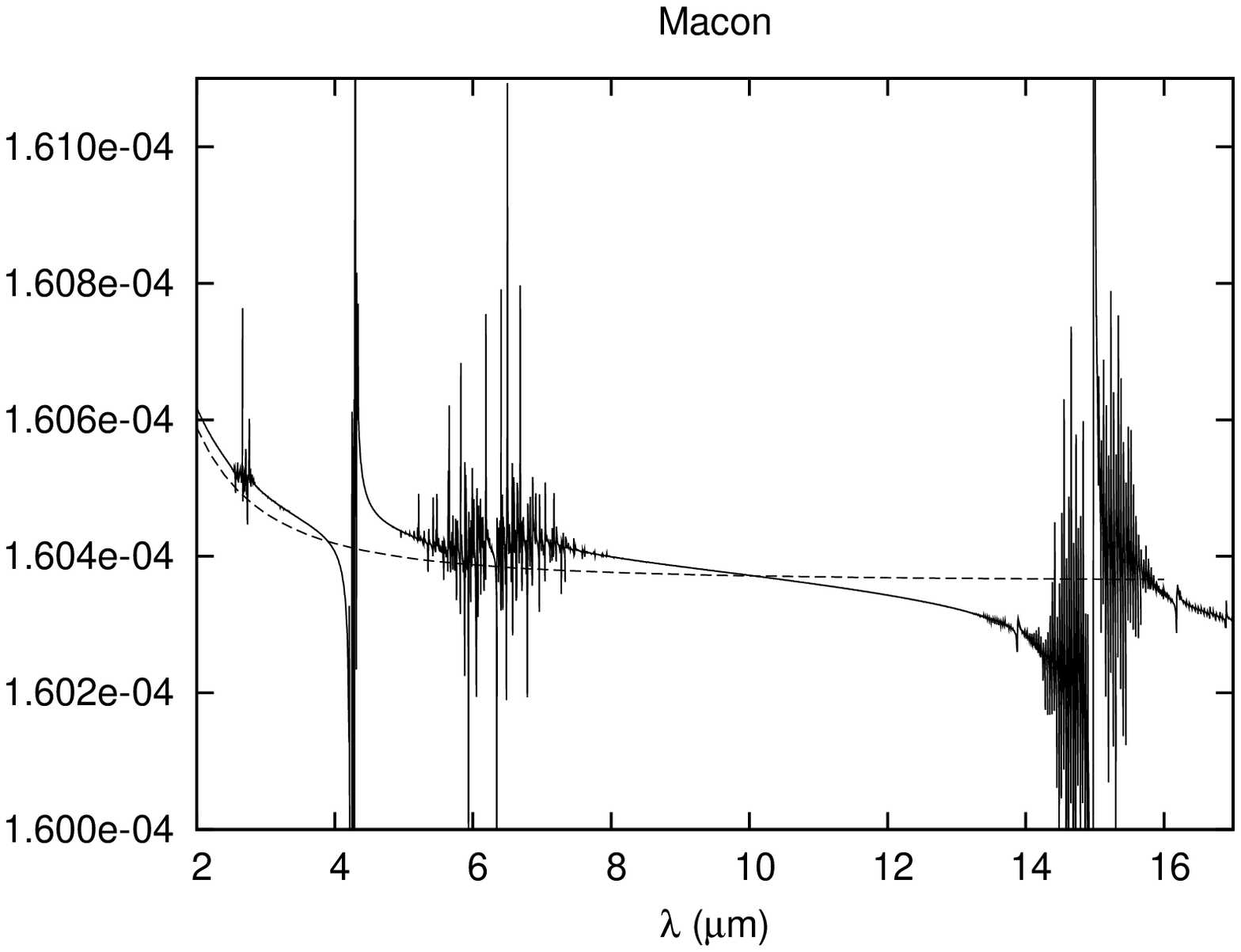}
\caption{Refractivity at the 2 representative site, Paranal (left) and Macon (right), using the parameters listed in Table~\ref{tab:atm_params}. The solid line follows Mathar, showing the water and CO$_{2}$ resonances. The dashed line shows the  extrapolation of the Ciddor formula to mid-IR wavelengths.}\label{fig:refractivity}
\end{figure}

In our subsequent discussion of atmospheric effects, we concentrate on atmospheric characteristics at two sites, Cerro Paranal (2400 m) and Cerro Macon (5000 m), representing a `standard' good site and a `high and dry site', respectively. Semi-empirical atmospheric profiles in temperature, pressure and water vapour mixing ratio were obtained in collaboration with ESO and the European Centre for Medium Range Weather Forecast\cite{profiles_ulli}. A summary of the relevant median parameters used in this paper for both sites is shown in Table~\ref{tab:atm_params}. The refractivity at these two sites is plotted in Fig.~\ref{fig:refractivity}, for both the Ciddor and Mathar formulations. The Mathar data show the location and effect on refractivity of the water and CO$_{2}$ resonances. The absolute water content at the `high and dry' site is smaller, causing a smaller slope (dispersion) of the refractivity. The relative humidity for the Macon parameters is nevertheless higher, 14.1 \%, compared to Paranal's 10.1 \%, because the water vapour saturation curve drops significantly over the 20 $^\circ$C difference between the two characteristic temperatures.

\bigskip
\begin{table}[h]
\centering
\begin{tabular}{|l|c|c|c|c|}
\hline
Site & h (m) & T (K) & P (mbar) & W.v. mixing ratio (g/kg)\\
\hline
Paranal & 2400 & 286.3 & 760.1 & 1.16\\
Macon & 5000 & 267.0 & 552.4 & 0.466\\
\hline
\end{tabular}
\caption{Median atmospheric parameters for the two representative sites, used as input for the calculations in this paper.}\label{tab:atm_params}
\end{table}

\section{Chromatic effects in adaptive optics correction}\label{sec:basic_chromatic}

In this section we discuss several problems expected to arise from atmospheric chromaticity, applied specifically to mid-infrared wavelengths. For refractivity calculations Mathar's theory is used to account for the resonances that are prominent in the IR refractivity curve of moist air (see Fig.~\ref{fig:refractivity}).

\subsection{Atmospheric dispersion}

The effect of simple atmospheric refraction in astronomical observations is to shift the apparent position of astronomical objects at non-zero zenith positions towards the zenith. This results in a `smearing' of the point spread function (PSF) along the zenith direction. The refraction, defined as the angular distance between apparent and true zenith distance, for a given refractive index $n$ is given by:

\begin{equation}\label{eq:R}
R  \approx 206,265(\frac{n^2-1}{2n^2})\tan(z)
\end{equation}
where $z$ is the true zenith distance in radians and $R$ is in arcseconds. The approximation assumes a spherically symmetric atmosphere and is valid for $z \le 80$\cite{astrophysical}. Full treatment should include the observatory altitude and latitude and target coordinates. From equation~\ref{eq:R} the atmospheric differential refraction between two wavelengths is thus given by:

\begin{equation}\label{eq:adr}
R_1-R_2=206,265(\frac{n_1^2-1}{2n_1^2}-\frac{n_2^2-1}{2n_2^2})\tan(z).
\end{equation}

\begin{figure}
\centering
\includegraphics[width=13cm]{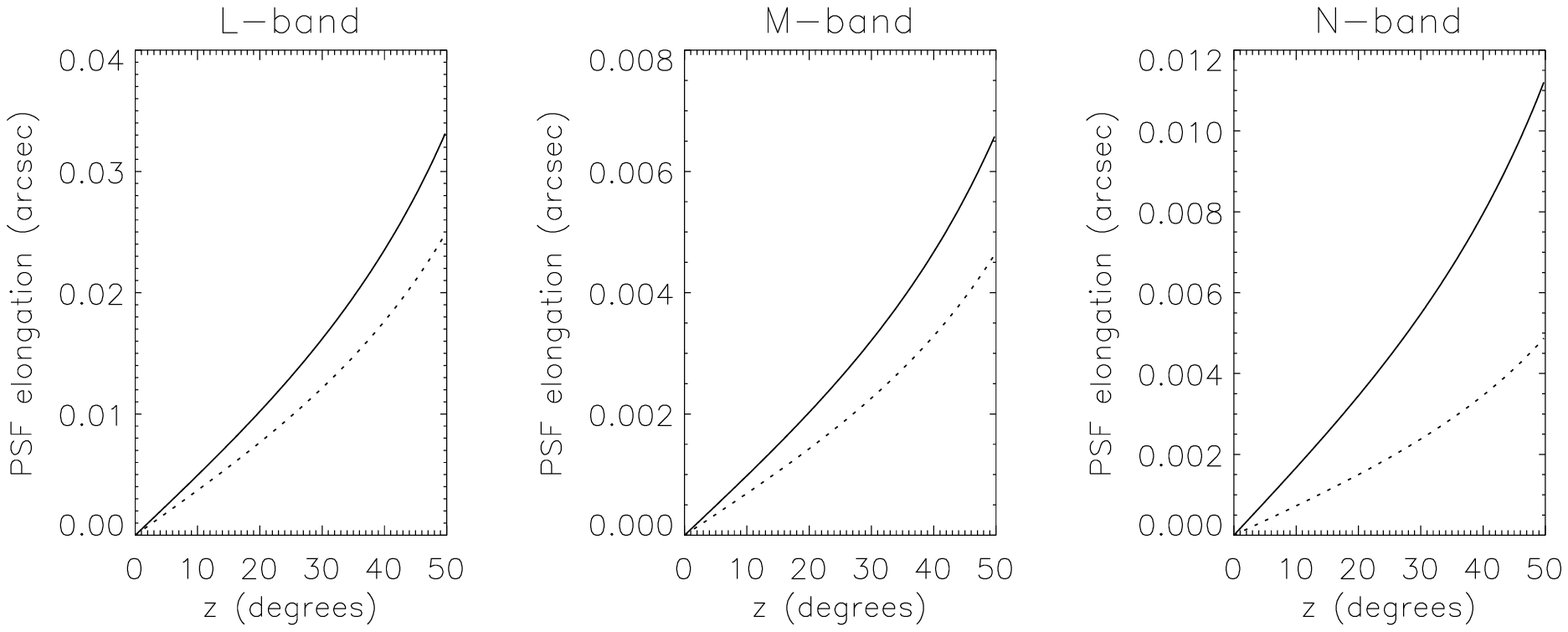}
\includegraphics[width=13cm]{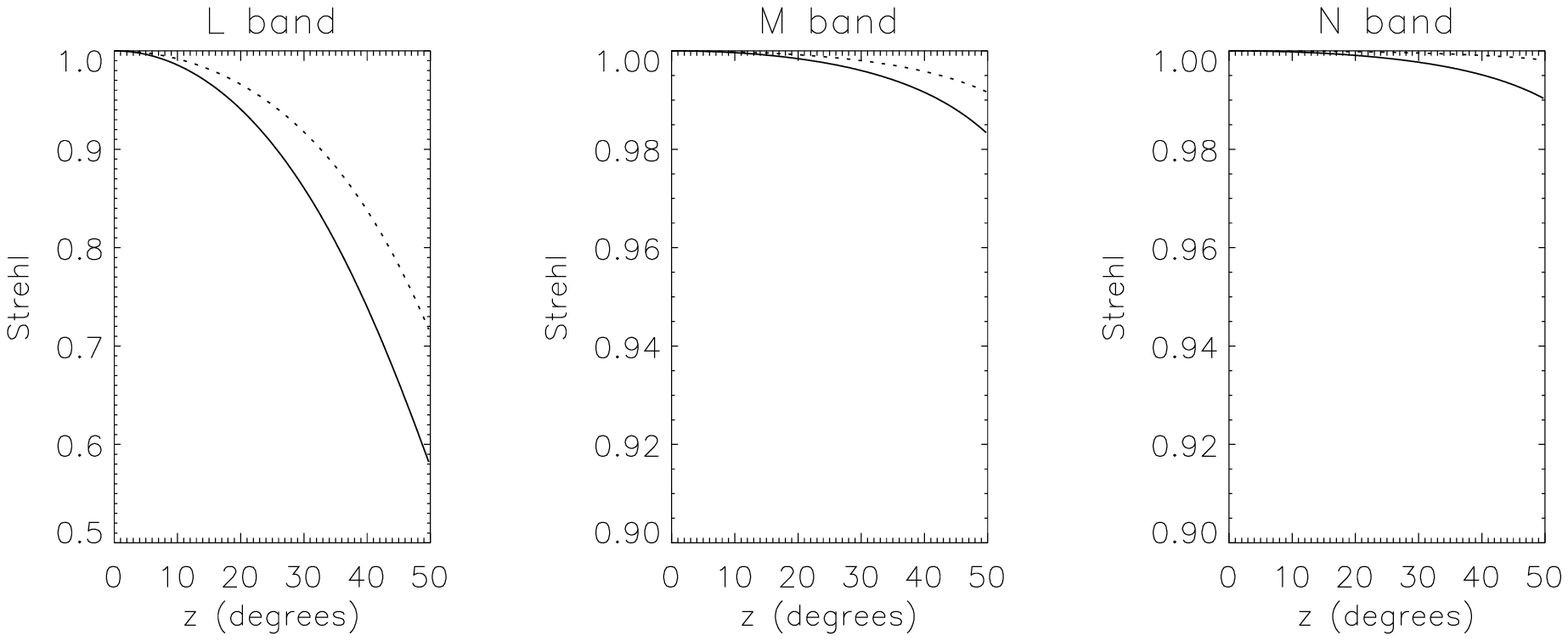}
\caption{Effect of atmospheric dispersion across the bandwidths of typical broad-band filters, as defined in Table~\ref{tab:filters}, as a function of zenith distance, around 3.8, 4.7 and 10.5 $\mu$m (L-, M- and N-bands) and for the 2 representative sites. Top row: dispersion across the bands in arcseconds. Bottom row: resultant Strehl loss from uncorrected dispersion. The solid line is for a regular site (Paranal), the dashed line for a high and dry site (Macon).}\label{fig:dispersion_bands}
\end{figure}

While the relative flatness of the dispersion curve at mid-IR wavelengths renders this effect largely irrelevant on today's 8-m telescopes, even in broad-band applications, the need for diffraction-limit operation on an ELT combined with the far superior diffraction limit of such telescopes, require revisiting the problem. Using the site parameters listed in Table~\ref{tab:atm_params}, we calculated refractivities for a representative broad-band filter in each of L-, M- and N-band, based on filters currently used on ground-based telescopes (see Table~\ref{tab:filters}). Equation~\ref{eq:adr} then yields the amount of elongation expected in the PSF of a broadband image at the E-ELT. The evolution of this PSF elongation with zenith distance is shown in Fig.~\ref{fig:dispersion_bands}.

The amount of elongation shown in this plot, of the order of 30-35 milli-arcsec(mas) in the worst case at z$=50^{\circ}$ at Paranal, represents almost twice the size of a diffraction-limited PSF core of a 42-m telescope in the L-band. The situation is markedly improved at a higher, drier site in all wavelength bands.

%
%
The expected Strehl loss from this level of dispersion, plotted in the bottom row of Fig.~\ref{fig:dispersion_bands},  was computed assuming that the source has a spectrum
that is flat over the specific bandwidth, that the dispersion is linear
over the bandwidth and that the diffraction limited source size is
approximately constant over the band width. The decrease can then be
computed by convolving the diffraction limited PSF by a top-hat function
with a width equal to the atmospheric dispersion and comparing the peak
intensity with the undispersed PSF. While for the M- and N-bands the resulting loss is minimal (of the order of $\sim$1-2\%), at shorter wavelengths its value can be several 10\% even at moderate z, with the value higher but still significantly reduced from 100\% at a high and dry site. Beyond imaging, for long slit spectroscopy dispersion will result in a distortion from wavelength-dependent light losses across the slit.

It would therefore seem that correction of atmospheric dispersion in its simplest form is required for diffraction-limited performance on an ELT, particularly at the shortest wavelengths (L-band). This is a very relevant conclusion given that no full-field compensator is foreseen in the baseline design of the E-ELT. Several schemes for full-field correction of atmospheric dispersion have been proposed in recent years~\cite{goncharov07, mette04,hawarden06}. However, correcting at mid-IR wavelengths using the same optics is not likely to be possible due to the limited transmission of the optics used.

With AO entirely integrated into the telescope's operation, a full treatment of refractive effects should include the dispersion as seen by the wavefront sensor (WFS), depending on WFS wavelength and bandwidth, and how it will affect the quality of the corrected image received by the instrument. This is discussed in the following section.

\begin{table}[h]
\centering
\begin{tabular}{|c|c|c|}
\hline
Band & $\lambda_{centre}$ ($\mu$m) & $\Delta\lambda$ ($\mu$m) \\
\hline
L & 3.78 & 0.70\\
M & 4.68 & 0.20\\
N & 10.5 & 1.50\\
\hline
\end{tabular}
\caption{Centre wavelengths and bandwidths for three representative filters used to calculate the effect of atmospheric dispersion in the mid-IR on the 42-m E-ELT.}\label{tab:filters}
\end{table}

\subsection{Refractivity and wavefront sensing}

Further degradation of the image quality due to differential refraction in the atmosphere is expected where WFS is carried out at a different wavelength to the AO correction. Several reasons exist for such a strategy:

\begin{itemize}
\item Use of monochromatic laser guide stars (LGS) at 589 nm;
\item Different colour of the natural guide star (NGS) to the science target;
\item If the science target is used as reference source, it may be better to conserve photons in the science band by sensing the wavefront at another wavelength .
\end{itemize}

We examine here two issues specifically related to wavefront sensing, when sensing wavelength differs from science wavelength.

\subsubsection{Time dependent differential refraction}

Because of the dispersion in the refractive index of air between visible and IR wavelengths, as shown in Fig.~\ref{fig:refractivity}, the pointing centres in the visible, in the WFS, and IR, at the science detector, are not coincident. While this instantaneous displacement is not necessarily a problem for an AO system, over long exposures with significant changes in zenith distances the offset will change. If the AO system holds fixed the visible pointing centre of the guide star, the IR image will gradually trail with respect to the science instrument, giving rise to a broadening of the PSF. If left uncorrected, this can impose a limitation on maximum exposure time. Roe~\cite{roe02} gives a comprehensive discussion of this effect, including the effect of observatory altitude and latitude and target coordinates, for a number of representative targets observed in H-band. His calculations show that the H-band PSF in a 30-m telescope at Mauna Kea is broadened by 25\% of the diffraction-limited core size within just one minute; he reports that the effect has already been observed on the 10-m Keck telescopes.

Potential solutions suggested by Roe are to include a continuous calculated correction into the AO correction, or inserting compensating optics between the telescope and AO system. In the case of the E-ELT, the latter option is unlikely to be feasible, due to the broad wavelength range to be corrected.

\begin{figure}[t]
\centering
(a)\includegraphics[width=6cm]{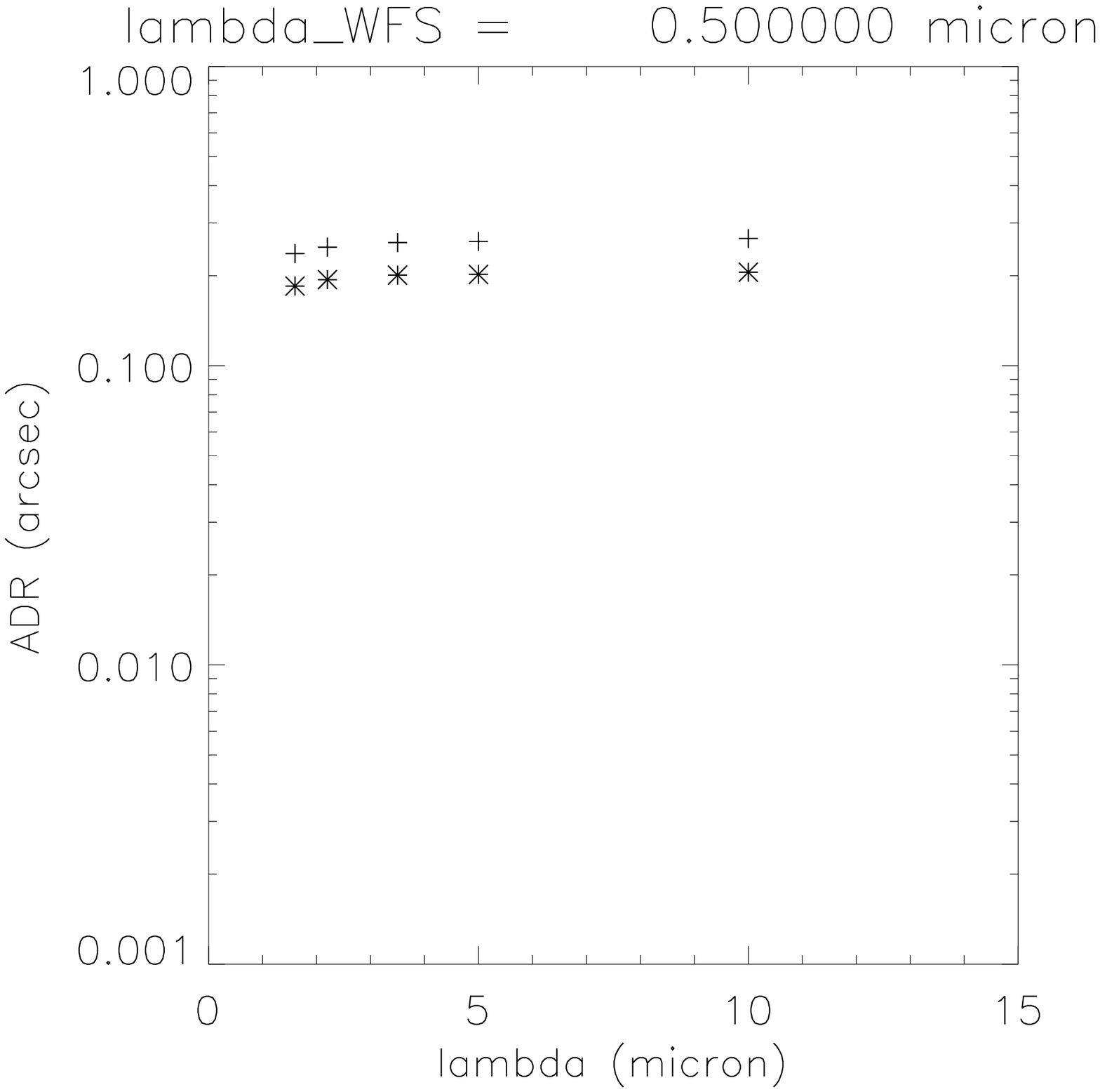}
(b)\includegraphics[width=6cm]{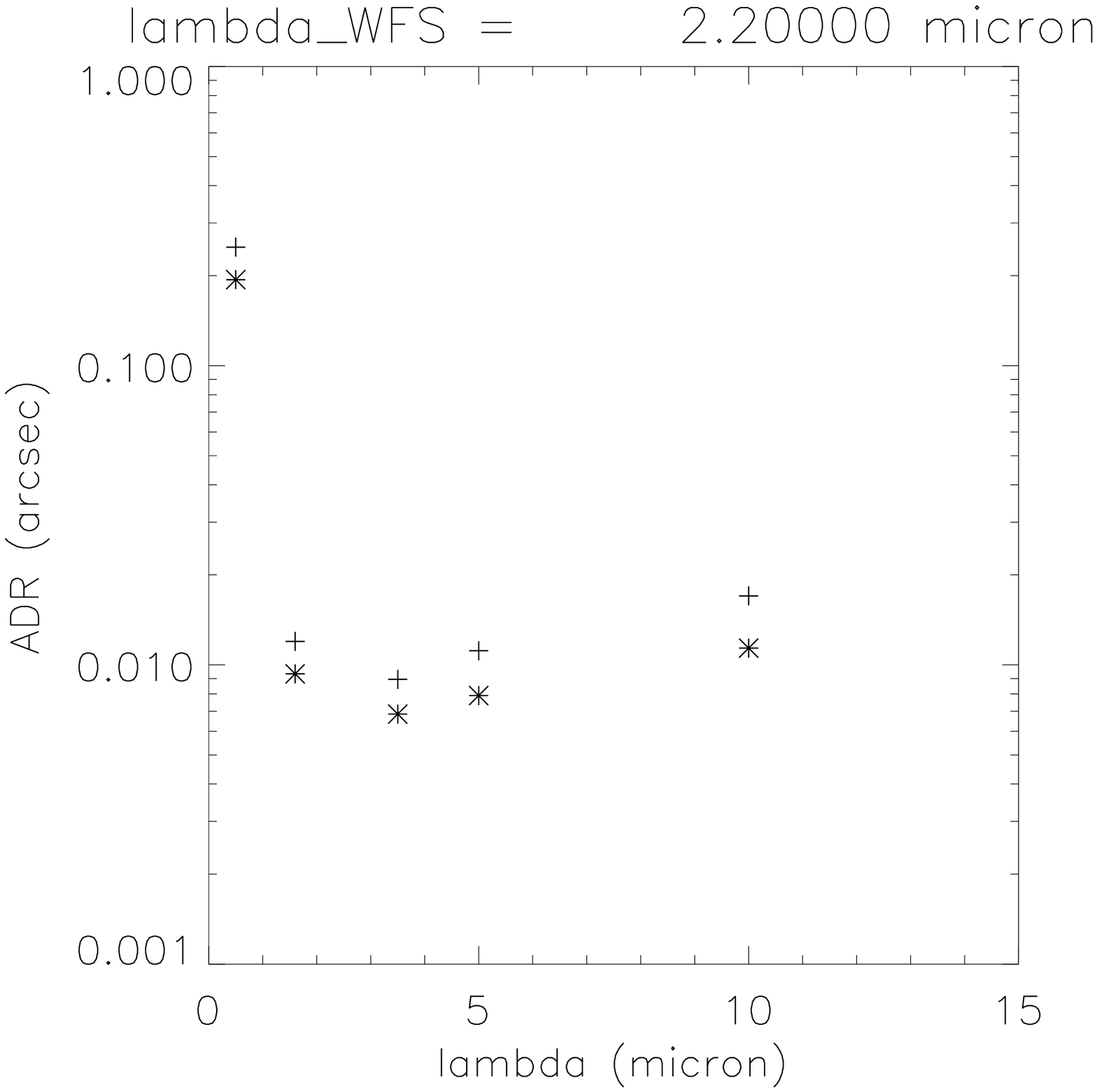}
(c)\includegraphics[width=6cm]{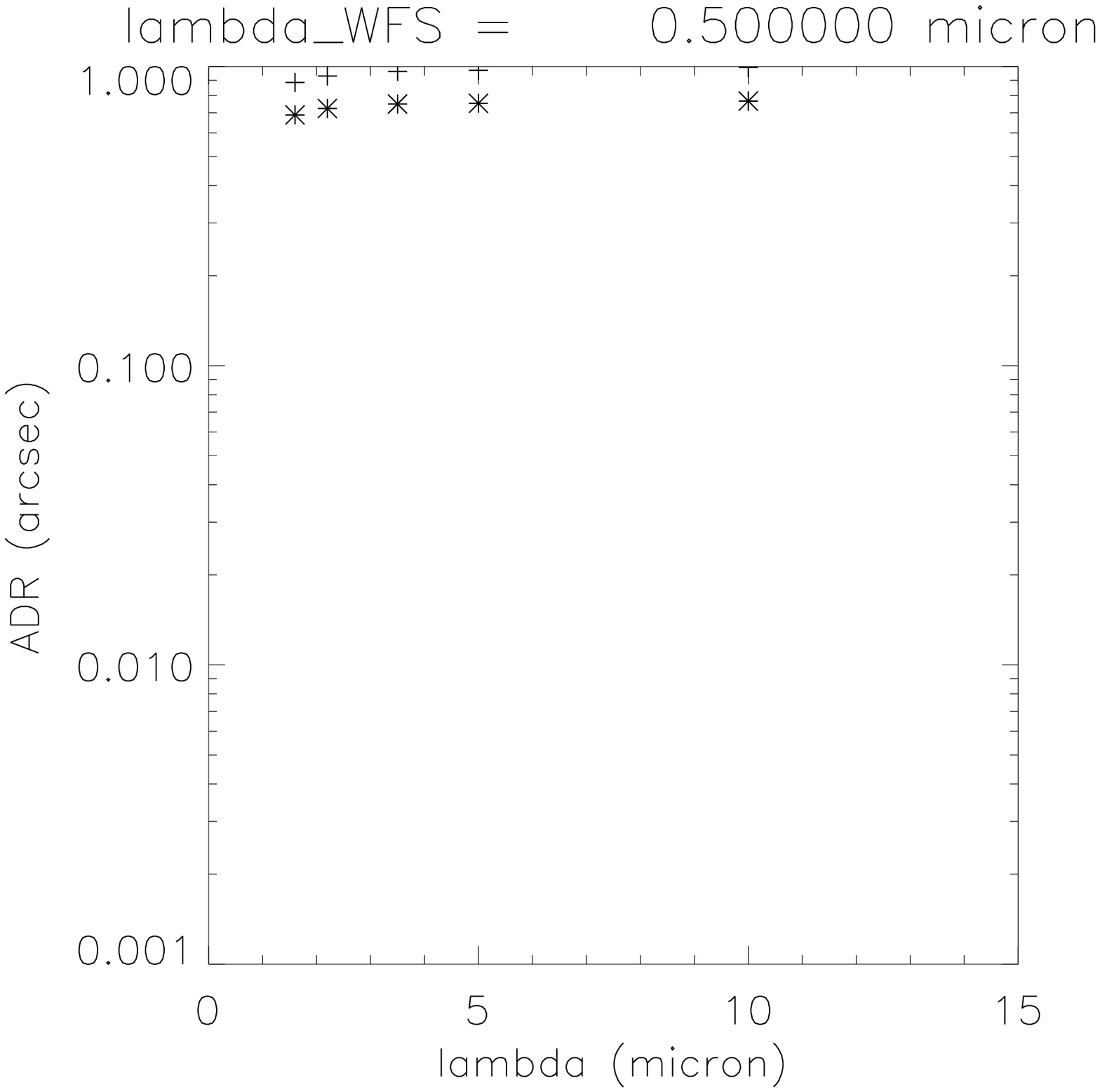}
(d)\includegraphics[width=6cm]{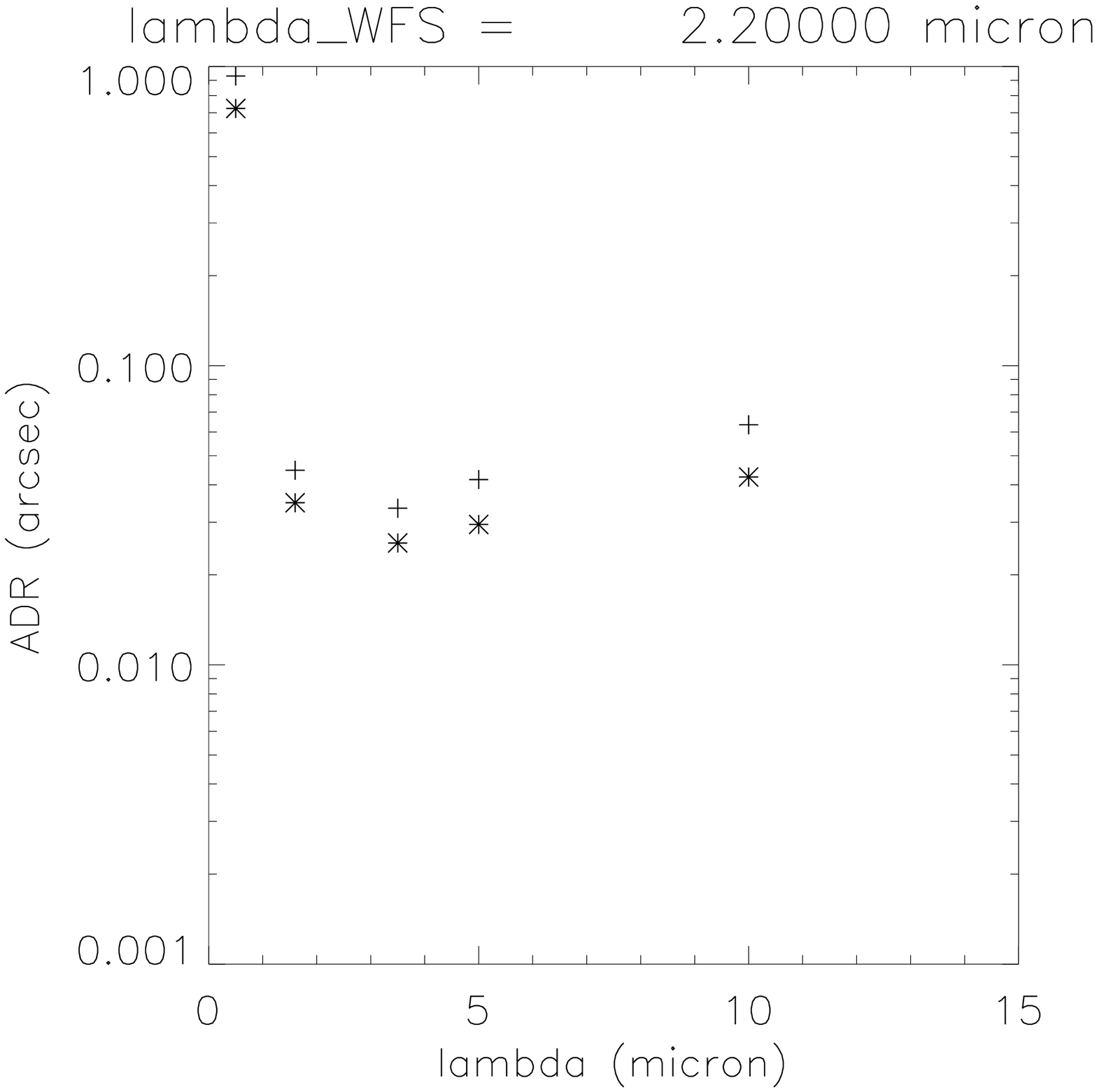}
\caption{Offset between visible and infrared pointing centres due to atmospheric differential refraction, calculated for representative wavelengths in V, H, J, K, L, M and N-bands. Top row: $z=15^{\circ}$, (a) $\lambda_{WFS}=0.5\mu m$, (b)  $\lambda_{WFS}=2.2\mu m$. Bottom row: $z=45^{\circ}$, (c) $\lambda_{WFS}=0.5\mu m$, (d) $\lambda_{WFS}=2.2\mu m$. Symbols: $+$ represents Paranal, $\ast$ Macon.}\label{fig:roecalcs}

\end{figure}

Using equations~\ref{eq:R} and~\ref{eq:adr} from the previous section, we can calculate the instantaneous offset between visible WFS and IR science image centres. The results, shown in Fig.~\ref{fig:roecalcs}, give a good indicator of the potential magnitude of the problem when observing at mid-IR wavelengths using AO for correcting atmospheric turbulence effects at both sites. When sensing is carried out in the V-band (left-hand panels), the plots show a displacement of pointing centres of the order of $250 mas$ at moderate $z$, and around $1 arcsec$ at $z=45^{\circ}$, at Paranal. Equivalent numbers at the high and dry site lie $\sim$ 20\% lower. Even without more detailed calculations, it is not unreasonable to expect that this will indeed place a limitation on exposure time of the magnitude described by Roe with a mid-IR ELT instrument. The right-hand panels in Fig.~\ref{fig:roecalcs}, showing the same data in the case of K-band wavefront sensing, seem to indicate that the problem will be much mitigated when wavefront sensing is carried out in the K-band (2.2 $\mu$m). This is expected, given the relative flattening of the refractivity curve beyond the NIR.

The above calculations assume the AO system makes use of a single natural guide star (NGS), and assume a monochromatic WFS. Telescopes in the ELT generation will however use advanced AO observing modes, such as ground-layer AO (GLAO) and laser tomography AO (LTAO), employing sodium laser guide stars (LGS) in asterism configurations around the science target. As these guide stars are from their very nature monochromatic at $\sim$600 nm, K-band WFS within the instrument will not be possible in these modes. Furthermore, LGS too will be subject to atmospheric refraction; detailed calculations following Roe's template extended to multiple guide star configurations will be needed to help develop a correction scheme.

These preliminary calculations indicate strongly that a solution for differential refraction between WFS and mid-IR science instrument is required for achieving near-diffraction-limited operation at these wavelengths on an ELT. For NGS AO, the problem can be solved to a large extent (though not entirely) by the inclusion of a K-band WFS in the instrument; although a full characterisation of the problem should include the effect of WFS bandwidth and spectral distribution of the NGS. When using LGS, however, WFS at NIR wavelengths is not possible because of their monochromatic nature. One solution would be to develop LGS at NIR wavelengths; we are unaware of such developments in the LGS community. In the absence of such a facility, a mid-IR AO-assisted instrument is likely to have to cope with Strehl losses where LGS observing modes are used. A detailed quantitative assessment is needed to further understand such a refractive effect.

\subsubsection{Chromatic amplitude sensing error}

A further problem arising from different sensing and correction wavelengths is the amplitude error on the WFS measurement. It can be shown that the wavefront can be written as the product of a chromatic function, which depends only on the wavelength, and another function which depends on the turbulent flow structure (temperature, pressure) only. In that case, it is clear that any given turbulent wavefront at a given wavelength can be translated to another wavelength, by simply using the dispersion formula; a wavefront measurement at the sensing wavelength should thus be easily transformable to the correction wavelength. However, this works perfectly only if both sensing and correction are monochromatic. In general, this is not the case, therefore a perfect compensation of this effect cannot be achieved. First order estimates suggest this to be a minor effect, giving rise to a Strehl loss of $<5 \%$ on an ELT at mid-IR wavelengths, and this effect will not be discussed in detail here.

\subsection{Chromatic anisoplanatism}

The dispersion of light of different colours causes them to travel through different paths in the atmosphere, and this gives rise to an anisoplanatic error. This phenomenon has been described in detail by several authors, for example Wallner~\cite{wallner84}, Nakajima~\cite{nakajima06} (who refers to it as `chromatic shear'). More recently, Devaney~\cite{devaney08} studied the effect in the context of ELTs and found that the physical separation of the rays, even between visible and NIR rays where dispersion is steeper, is small compared with the value of $r_0$ at these wavelengths, giving rise to Strehl losses of $< 1\%$ even for high z. We will therefore not provide a further discussion of this effect for mid-IR observations.

\section{Water vapour turbulence}\label{sec:wvturb}

As the air refractive index depends on the temperature, pressure, and air constituents, as well as the wavelength, the refractive index structure constant $C_{N}^{2}$ also depends on these parameters. Therefore, $C_{N}^{2}$ can be split into several subcomponents, one for each atmospheric parameter plus cross-couplings. And each of these subcomponents depends on the wavelength as well. Hill et al.\cite{hill80} argue convincingly that the only parameters with a significant impact on the $C_{N}^{2}$ model are fluctuations in temperature and humidity, with contributions from pressure variations and those arising from other minor constituents, such as CO$_2$ negligible. We therefore get:

\begin{equation}
C_{N}^{2}=A^{2}_{T}\,\frac{C_{T}^{2}}{\left<T\right>^{2}}+
          A^{2}_{Q}\,\frac{C_{Q}^{2}}{\left<Q\right>^{2}}+
          2\,A_{T}\,A_{Q}\,\frac{C_{T,Q}}{\left<T\right>\left<Q\right>}
\end{equation}
where $A_{T}$ and $A_{Q}$ are functions of the pressure, temperature, water vapor pressure, and wavelength. Interestingly, the cross-coupling can have either positive or negative values, in principle, therefore it is possible that in certain conditions the presence of water vapour compensates partially the temperature-based $C_{N}^{2}$. In normal conditions, and in the visible and near-IR, optical turbulence due to temperature fluctuation largely dominates the water-based optical turbulence\cite{roddier81}.

At far-IR and submillimetre wavelengths, and in the interferometry community, the effect of water vapour turbulence is well known to affect the quality of observations. Indeed, phase variance due to atmospheric water vapour fluctuations is an important problem for many present and future facilities, such as the Very Large Telescope Interferometer (VLTI)\cite{mathar_paranal07}, the Keck interferometer\cite{colavita04}, and the Atacama Large Millimetre Array (ALMA)\cite{nikolic08}. The available experimental data largely employs water vapour radiometry or observations of radio beacons from geostationary satellites, and have yielded useful results. Measurements at the ALMA site~\cite{alma_517, alma_543} have shown a Kolmogorov-like power spectrum of water vapour turbulence, with a night-time vertical scale height of approximately 1 km. However, there is a need for more experimental data, particularly relevant to mid-IR observations with an ELT-sized filled aperture, and verification of the atmospheric models used for interpretation of data.

Hill's formula permits the calculation of the $C_{N}^{2}$, as the functions $A_{T}$ and $A_{Q}$ can be computed from a knowledge of the temperature, pressure and humidity vertical profiles, but the $C_{T}^{2}$, $C_{Q}^{2}$ and $C_{T}C_{Q}$ need to be measured on-site. As reliable experimental data are lacking, we perform a first-order estimation of the impact of the additional water vapour-based turbulence.
A rough estimate of the proportion of water vapour and dry air (temperature) optical turbulence can be found in Colavita et al.~\cite{colavita04}: by extrapolating water vapour turbulence data measured by radio interferometers at Mauna Kea (measuring rms phase delay fluctuation over a 100 m baseline) to the visible and mid-IR, and assuming an outer scale (L$_0$) of 40 m, they estimate that the refractive index fluctuation due to water vapor would be about 1/20th of the temperature-based refractive index fluctuation, in the visible and near-IR, and about 1/7th in the mid-IR at 10 microns. Translated into a ratio of $C_{N}^{2}$, these numbers would be respectively 1/400th and 1/49th.

We consider the case where WFS is carried out at a wavelength not sensitive to the presence of water vapour (i.e. visible or NIR), correction is applied at 10 $\mu$m (N-band), and the water vapour turbulence is statistically decorrelated from traditional temperature-based turbulence (i.e. neglecting any cross-coupling). The result is an independent wavefront error in the corrected image that remains uncorrected, as it can not been `seen' at the sensing wavelength. We can thus write the phase variance associated to this unseen wavefront error using the classical expression for uncorrected turbulent phase error:

\begin{equation}
\sigma_{WV}^{2}=1.0324\,\eta(L_0/D)\,(D/r_{0})^{5/3}
\end{equation}
where the Fried parameter is calculated from $C_{N}^{2}$ using
 
\begin{equation} 
r_{0}^{-5/3}=0.4234(2\pi/\lambda)^{2}\int C_{N}^{2}(h)dh,
\end{equation} 

L$_0$ is the outer scale, D the aperture diameter, and $\eta(L_0/D)$ is the outer scale attenuation factor (see Jolissaint et al.~\cite{jolissaint04} for an accurate empirical calculation method).

In Table~\ref{tab:watervaporturb}, we show the N-band (10 $\mu$m) Strehl ratio associated to the variance error above for three values of the outer scale L$_0$ (infinite, 60 m and 30 m), three telescope diameters (8, 30 and 42 m), a dry turbulence characterized by a typical $r_{0}$ of 10 cm, and assuming Colavita's values for the refractive index fluctuations ratio (water/temperature). It is found that for large apertures, the Strehl loss can become quite significant, if the outer scale is larger than the typical values encountered at most locations (in the range 20-40 m). For 8-m telescopes the effect is shown to be very small. In the absence of dedicated measurements of the water vapour optical turbulence in the mid-IR, it is difficult to assess the importance of this effect. Our estimation at least shows that it is not negligible, and can become a serious issue for large apertures; given the lack of understanding of water vapour turbulence phenomena, these results should be viewed with caution. Water vapour optical turbulence measurement campaigns, on baselines relevant to observations with an ELT-sized filled aperture, are critically needed to support mid-IR AO performance analysis in the context of the ELT projects.

\begin{table}[htdp]
\centering
\begin{tabular}{|l|c|c|c|}
\hline $L_{0}$  & 8 m  & 30 m & 42 m\\
\hline $\infty$ & 0.93 & 0.50 & 0.30\\
60 m     & 0.97 & 0.90 & 0.88\\
30 m     & 0.98 & 0.96 & 0.95\\
\hline
\end{tabular}
\caption{Estimated N-band Strehl loss in the presence of (uncorrected) water vapour turbulence, based on measurement data from Mauna Kea extrapolated to mid-IR wavelengths.}\label{tab:watervaporturb}
\end{table}

\section{Conclusions}

Atmospheric dispersion is a well known source of error in astronomical observations at visible and near-infrared wavelengths. With the advent of ELTs and their integrated AO facilities, atmospheric refractivity becomes a relevant quantity even in the mid-infrared. This is particularly the case if the traditional refractivity formulations are recalculated to include the effects of IR water and CO$_2$ resonances, which causes the refractivity to deviate substantially from the NIR values extrapolated to longer wavelengths.

The calculations in this paper, though in many places approximate in nature, show the likely magnitude of refractive effects on the AO correction for a mid-IR ELT instrument. While `simple' dispersion will only cause a significant drop in Strehl at the short-wavelength end of the region, differential refraction effects between visible/NIR WFS and mid-IR science instrument may cause a significant image degradation over long exposures, even at moderate zenith distances. This problem in particular will be investigated more closely, taking into account the effect of multiple laser guide stars which can only be sensed in the visisble. 

Water vapour turbulence has recently been of interest in the interferometry and submillimetre communities, however, a thorough understanding of the phenomenon and experimental data are lacking (and non-existent in the framework of an ELT-sized filled aperture). Our first order estimates show that the Strehl loss can be significant, if not large, depending on the size of the outer scale. However, given the poor understanding of this type of turbulence both in the spatial and temporal domain, and of its correlation with dry air turbulence, these results show be regarded with caution. Further experimental results, specifically designed for ELT-type observations in the mid-IR, are required to help assess the problem.

The comparison of results between the a Paranal-like site and a `high and dry' location like Cerro Macon is also very important, and reinforces the notion that a high and dry location is more suitable for mid-IR astronomy. However, we expect some of these advantages to apply to NIR also, and would invite NIR colleagues to examine site-related effects more closely.

Correcting atmospheric dispersion on an ELT is a challenging task, even in today's telescopes. Solutions have been suggested for ELTs in recent years, but none have the capability of providing correction over a wide wavelength range (from visible to infrared). The problems will have to be addressed by a combination of technology (e.g. K-band wavefront sensing, NIR laser guide stars) and careful calibration. The first step, however, is a detailed theoretical understanding, and we intend to continue this work to make this possible.

\section{Acknowledgements}

The authors wish to thank John Richer, Henry Roe and Jacques Beckers for informative discussions, and exchange of software routines, papers and data. We are also grateful to the METIS consortium for supporting and participating in this work.

\bibliography{midir_bib}

\begin{thebibliography}{10}

\bibitem{gilmozzi07}
R.~{Gilmozzi} and J.~{Spyromilio}, ``{The European Extremely Large Telescope
  (E-ELT)},'' {\em The Messenger}~{\bf 127}, p.~11, Mar. 2007.

\bibitem{crampton07}
D.~{Crampton}, L.~{Simard}, and D.~{Silva}, ``{TMT Science and Instruments},''
  in {\em {Proceedings of the ESO workshop on Science with VLT in the ELT
  era}},  2007.

\bibitem{brandl_metis08}
B.~{Brandl} {\em et~al.}, ``{METIS},'' in {\em {Proc. SPIE vol. 7014,
  'Ground-based and airborne instrumentation for astronomy II'}},  I.~S.
  {MacLean} and M.~{Casali}, eds., 2008.

\bibitem{astrophysical}
A.~N. {Cox}, ed., {\em {Allen's Astrophysical Quantities}}, {AIP
  Press/Springer}, 4th~ed., 2000.

\bibitem{filippenko82}
A.~V. {Filippenko}, ``{The importance of atmospheric differential refration in
  spectrophotometry},'' {\em {PASP}}~{\bf 94}, pp.~715--721, 1982.

\bibitem{reardon06}
K.~P. {Reardon}, ``{The effects of atmospheric dispersion on high-resolution
  solar spectroscopy},'' {\em {Solar Physics}}~{\bf 239}, pp.~503--517, 2006.

\bibitem{arribas99}
S.~{Arribas}, E.~{Mediavilla}, B.~{Garcia-Lorenzo}, C.~{del Burgo}, and J.~J.
  {Fuensalida}, ``{Differential atmospheric refraction in integral field
  spectroscopy: Effects and correction},'' {\em {Astron. Astrophys. Suppl.
  Ser.}}~{\bf 136}, pp.~189--192, 1999.

\bibitem{mette06}
M.~{Owner-Petersen}, ``{Effects of atmospheric dispersion on the PSF background
  level},'' in {\em {Proc. SPIE vol. 6272}},  2006.

\bibitem{roe02}
H.~G. {Roe}, ``{Implications of atmospheric differential refration for adaptive
  optics observations},'' {\em {PASP}}~{\bf 114}, pp.~450--461, 2002.

\bibitem{helminiak08}
K.~G. {Helminiak}, ``{Impact of the atmospheric refraction on the precise
  astrometry with adaptive optics in the infrared},'' {\em arXiv:0805.3369v1
  [astro-ph]} , 2008.

\bibitem{hawarden06}
T.~G. {Hawarden}, E.~{Atad-Ettedgui}, C.~R. {Cunningham}, D.~M. {Henry}, C.~J.
  {Norrie}, M.~{Wells}, J.~C. {Dainty}, N.~{Devaney}, and A.~V. {Goncharov},
  ``{European Extremely Large Telescope Design Study: Atmospheric dispersion
  correction. Final report.},'' tech. rep., {UKATC, NUI Galway}, 2006.

\bibitem{hitran}
L.~S. {Rothman} {\em et~al.}, ``The {HITRAN} 2004 molecular spectroscopic
  database,'' {\em J.\ Quant.\ Spectrosc.\ Radiat.\ Transfer}~{\bf 96},
  pp.~139--204, 2005.

\bibitem{hill80}
R.~J. {Hill}, S.~F. {Clifford}, and R.~S. {Lawrence}, ``{Refractive index and
  absorption fluctuations in the infrared caused by temperature, humidity, and
  pressure fluctuations},'' {\em {J. Opt.\ Soc.\ Am.}}~{\bf 70},
  pp.~1192--1205, 1980.

\bibitem{hill86}
R.~J. {Hill} and R.~S. {Lawrence}, ``{Refractive index of water vapor in
  infrared windows},'' {\em {Infrared Physics}}~{\bf 26}, pp.~371--376, 1986.

\bibitem{mathar_ao04}
R.~J. {Mathar}, ``{Calculated Refractivity of Water Vapor and Moist Air in the
  Atmospheric Window at 10 $\mu$m},'' {\em {Appl.\ Op.}}~{\bf 43}(4),
  pp.~928--932, 2004.

\bibitem{mathar07a}
R.~J. {Mathar}, ``{Refractive index of humid air in the infrared: Model
  fits},'' {\em J. Opt. A.: Pure Appl. Opt.}~{\bf 9}(5), pp.~470--476, 2007.

\bibitem{edlen53}
B.~{Edl\'en}, ``{The dispersion of standard air},'' {\em {J. Opt.\ Soc.\
  Am.}}~{\bf 43}(5), pp.~339--344, 1953.

\bibitem{edlen66}
B.~{Edl\'en}, ``{The refractive index of air},'' {\em {Metrologia}}~{\bf 2}(2),
  pp.~71--81, 1966.

\bibitem{birchdowns93}
K.~{Birch} and M.~{Downs}, ``An updated edlen equation for the refractive index
  of air,'' {\em Metrologia}~{\bf 30}, pp.~155--162, 1993.

\bibitem{birchdowns94}
K.~{Birch} and M.~{Downs}, ``{Correction to the updated Edlen equation for the
  refractive index of air},'' {\em Metrologia}~{\bf 31}, pp.~315--316, 1994.

\bibitem{ciddor96}
P.~E. {Ciddor}, ``{Refractive index of air: New equations for the visible and
  near-infrared},'' {\em {Appl. Opt.}}~{\bf 35}(9), pp.~1566--1573, 1996.

\bibitem{bonsch98}
G.~{B\"onsch} and E.~{Potulski}, ``{Measurement of the refractive index of air
  and comparison with modified Edl\'en's formulae},'' {\em {Metrologia}}~{\bf
  35}, pp.~133--139, 1998.

\bibitem{colavita04}
M.~M. {Colavita}, M.~R. {Swain}, R.~{Akeson}, C.~D. {Koresko}, and R.~J.
  {Hill}, ``{Effects of atmospheric water vapor on infrared interferometry},''
  {\em PASP}~{\bf 116}, pp.~876--885, 2004.

\bibitem{profiles_ulli}
U.~{K\"{a}ufl}, ``{Semi-empirical atmospheric profiles}.'' Personal
  communication, Nov 2007.

\bibitem{goncharov07}
A.~V. {Goncharov}, N.~{Devaney}, and J.~C. {Dainty}, ``{Atmospheric dispersion
  compensation for extremely large telescopes},'' {\em {Optics Express}}~{\bf
  15}(4), pp.~1534--1542, 2007.

\bibitem{mette04}
M.~{Owner-Petersen}, ``{Some consequences of atmospheric dispersion for
  ELTs},'' in {\em {Proc. SPIE vol. 5489, Ground-based telescopes}},  J.~M.
  {Oschmann}, ed., pp.~507--517, 2004.

\bibitem{wallner84}
E.~P. {Wallner}, ``{Comparison of diffractive and refractive effects in
  two-wavelength adaptive transmission},'' {\em {JOSA A}}~{\bf 1}(7),
  pp.~785--787, 1984.

\bibitem{nakajima06}
T.~{Nakajima}, ``{Zenith-distance dependence of chromatic shear effect: A
  limiting factor for an extreme adaptive optics system},'' {\em {ApJ}}~{\bf
  652}, pp.~1782--1786, 2006.

\bibitem{devaney08}
N.~{Devaney}, A.~V. {Goncharov}, and J.~C. {Dainty}, ``{Chromatic effects of
  the atmosphere on astronomical adaptive optics},'' {\em {App. Op.}}~{\bf 47},
  pp.~1072--1081, 2008.

\bibitem{roddier81}
F.~{Roddier}, {\em {Progress in Optics XIX}}, ch.~{The effects of atmospheric
  turbulence in optical astronomy}, pp.~283--376.
\newblock {North-Holland}, 1981.

\bibitem{mathar_paranal07}
R.~J. {Mathar}, ``{High frequency fluctuations of water and carbon dioxide,
  Paranal June 25-30, 2007},'' tech. rep., {University of Leiden}, 2007.

\bibitem{nikolic08}
B.~{Nikolic}, J.~{Richer}, R.~{Hills}, and A.~{Stirling}, ``{Phase correction
  for ALMA: Adaptive optics in the submillimetre},'' {\em {The Messenger}}~{\bf
  131}, pp.~14--19, 2008.

\bibitem{alma_517}
A.~{Stirling}, J.~{Richer}, R.~{Hills}, and A.~{Lock}, ``{Turbulence
  simulations of dry and wet phase fluctuations at Chajnantor. Part I: The
  daytime convective boundary layer},'' tech. rep., ALMA Memo no. 517, 2005.

\bibitem{alma_543}
Y.~{Robson}, ``{Phase fluctuations at the ALMA site and height of the turbulent
  layer},'' tech. rep., {ALMA Memo no. 543}, 2001.

\bibitem{jolissaint04}
L.~{Jolissaint}, J.-P. {V\'eran}, and J.~{Marino}, ``{OPERA, an automatic PSF
  reconstruction software for Shack-Hartmann AO systems: Application to
  ALTAIR},'' in {\em {Proc. SPIE vol. 5490}},  pp.~151--163, 2004.

\end{thebibliography}
\bibliographystyle{spiebib}

\end{document}